\documentclass[showpacs,preprintnumbers,aps,prl,twocolumn,floatfix]{revtex4}

\usepackage[dvips,xdvi]{graphicx}
\usepackage{epsfig}
\usepackage{amsmath}

\begin{document}
\title{Evanescent quadrupole polariton}
\author{Oleksiy Roslyak and Joseph L. Birman}
\affiliation{Physics Department, The City College, CUNY\\
Convent Ave. at 138 St, New York, N.Y. 10031, 
USA}
\date{\today}
\preprint{APS/PREPRINT}

\begin{abstract}
In the work we demonstrate the formation of new type of polariton on the interface between a cuprous oxide slab and a polystyrene micro-sphere placed on the slab. The evanescent field of the resonant whispering gallery mode (WGM) has a substantial gradient, and therefore effectively couples with the quadrupole $1S$ excitons in cuprous oxide. This evanescent polariton has a long life-time ($1.7 ns$), which is determined only by its excitonic component. The polariton lower branch has a well pronounced minimum. This suggests that this excitation can be utilized for BEC. The spatial coherence of the polariton can be improved by assembling the micro-spheres in a linear chain.
\end{abstract}
\maketitle
\section{Introduction}
Although quadrupole excitons in cuprous oxide crystals are good candidates for BEC due to their narrow line-width and long life-time there are some factors impeding BEC \cite{KAVOULAKIS:1996, ROSLYAK:2007}. One of these factors is that due to small but non negligible coupling to the photon bath, one must consider BEC of the corresponding mixed light-matter states called polaritons \cite{FROHLICH:2005}. The photon-like part of the polariton has a large group velocity and tends to escape from the crystal. Thus, the temporal coherence of the condensate is effectively broken \cite{ELL:1998}. One proposed solution to this issue is to place the crystal into a planar micro-cavity \cite{KASPRZAK:2006}. But even state-of-the-art planar micro-cavities can hold the light no longer than $10 \ \mu s$. 
\par
Therefore in this work we propose to impede the polariton escaping by trapping it into a whispering gallery mode (WGM) of a polystyrene micro-sphere (PMS).
\section{the model}
We assume that the PMS of radius $r_0 \ \mu m$ is placed at a small \footnote{comparing to the evanescent field penetration depth } distance $\delta r_0 \ll r_0$ from the cuprous oxide crystal ($\epsilon_{Cu2O} =6.5$). Some density of quadrupole $1S$ excitons ([QE], $\hbar \omega_{1S} = 2.05 \ eV$, $\lambda_{1S} = 2 \pi / \omega_{1S} = 6096 \ \AA$) has been created by an external laser pulse. The corresponding polaritons move in the crystal by diffusion and can be trapped at the surface by the micro-sphere. 
\par
Because the evanescent field penetration depth ($\lambda_{1S}/2 \pi \left({\epsilon_{Cu2O}-1}\right)^{1/2} = 414 \ \AA$)  is much bigger than the QE radius ($a_B = 4.6 \ \AA$) the light-matter interaction can be considered semi-classically. For resonance coupling with a WGM its size parameter should be determined by the resonant wave vector in the cuprous oxide $k_0 = 2.62 \times 10^7 \ m^{-1}$. For example, if one takes a polystyrene ($\epsilon^2=1.59$) sphere of radius $r_0 = 10.7 \ \mu m$ then $k_0 r_0 = 28.78350$ and this corresponds to the 39TE1 resonance \cite{MIYAZAKI:2000}.  
\par
The light part of the polariton trapped inside the PMS moves as it would move in a micro-cavity of the effective modal volume $V \ll 4 \pi r_0^3 /3$. Consequently, it can escape through the evanescent field. This evanescent field essentially has a quantum origin and is due to tunneling through the potential caused by dielectric mismatch on the PMS surface. Therefore, we define the \emph{evanescent} polariton (EP) as an evanescent light-matter coherent superposition.
\par
The evanescent light has small intensity. Therefore it is not effective for the dipole allowed coupling. But it has a large gradient, so it can effectively couple through its quadrupole part.
\par
Let us assume that the incident polariton wave vector is along the interface and runs in the $z$ direction, the polarization of the polariton is along $x$ direction. Therefore, in the system of coordinates centered at the sphere, the incident polariton light part can be written as \cite{BOHREN:1983}:
\begin{equation}
\label{EQ:1}
{\bf E}_i = E_0 i^l \frac{2l+1}{l\left({l+1}\right)}\left({{\bf M}_{1l} - i {\bf N}_{1l}}\right),
\end{equation}
where ${\bf M}_{1l}$ and ${\bf N}_{1l}$ are vector spherical harmonics corresponding to TE- and TM-polarized modes of angular momentum $l$ and $z$ component of the angular momentum is $\left|m\right|=1$; $E_0$ is the amplitude of the electric field. The scattered field is given as:
\begin{equation*}
{\bf E}_s = E_0 i^l \frac{2l+1}{l\left({l+1}\right)}\left(i{a_{1l}{\bf N}_{1l} - b_{1l}{\bf M}_{1l}}\right),
\end{equation*}
where $a_{1l}$ and $b_{1l}$ are scattering Mie coefficients (See the Appendix). Keeping only the resonant term the last expression yields:
\begin{equation}
\label{EQ:2}
{\bf E}_s = -E_0 i^l 0.05 b_{1,39}{\bf M}_{1,39},
\end{equation}
While in the system of the coordinate, centered at the cuprous oxide, the plane wave is still given by the expression (\ref{EQ:1}), the scattered field has to be changed according to the vector spherical harmonic addition theorem \cite{STEIN:1961}:
\begin{equation}
\label{EQ:3}
{\bf M}_{1,39}=A^{ml}_{1,39}\left({r_0 + \delta r}\right){\bf M}_{ml}+B^{ml}_{1,39}\left({r_0 + \delta r}\right){\bf N}_{ml}
\end{equation}
Here $A^{ml}_{1,39}$ and $B^{ml}_{1,39}$ are the translational coefficients. Their explicit expression can be found, for instance, in \cite{FULLER:1991,MIYAZAKI:2000} and are explicitly listed in the Appendix.
\par     
The bulk (incident) and evanescent polaritons in cuprous oxide are formed through the quadrupole part of the light-matter interaction:
\begin{equation*}
H_{int} = \frac{i e }{m \omega_{1S}} {\bf E}_{i,s} \cdot {\bf p}
\end{equation*}
Here $e,m$ are the electron mass and charge, and ${\bf p}$ is the electron momentum. For the quadrupole $1S$ transition in cuprous oxide the energy of the interaction can be written as:
\begin{equation}
\label{EQ:4}
g=\left\langle{^3\Gamma^+_{5,xz}}\left| {H_{int}} \right|{^1\Gamma^+_1}\right\rangle = \left\langle{^3\Gamma^+_{5;1,2}}\left| {H_{int}} \right|{^1\Gamma^+_{1;0,0}}\right\rangle
\end{equation}
Here we introduced the initial state of the system, which transformes as irreducible representation $^1\Gamma^+_1$ of the cubic centered group $O_h$. The final state is the ortho-exciton state which transforms as $^3\Gamma^+_{5,xz}$ in Cartesian system or as $^3\Gamma^+_{5;1,2}$ in the corresponding spherical basis.
\par
\begin{figure}[htbp]
	\centering
		\includegraphics[width=8.6cm]{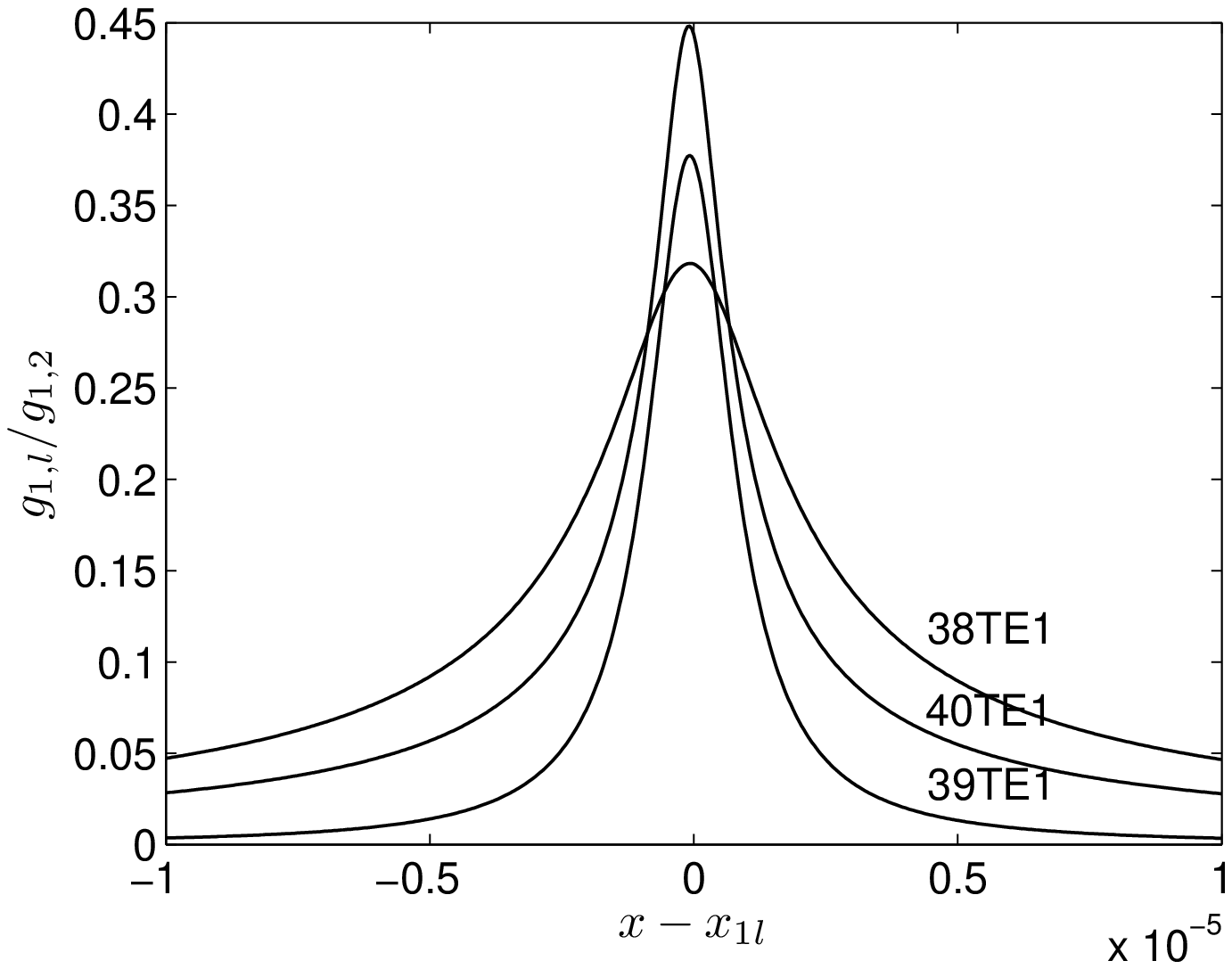}
	\caption{The evanescent light - $1S$ quadrupole coupling ($g_{1,l}$) scaled to the bulk exciton-photon coupling ($g_{1,2}$). The size parameter $kr_0$ is denoted as $x$ and the PMS is placed directly on the cuprous oxide sample ($\delta r =0$, See also Fig.\ref{FIG:2}).}
	\label{FIG:1}
\end{figure}

Hence, using (\ref{EQ:1}, \ref{EQ:2}, \ref{EQ:3}, \ref{EQ:4}), one can deduce that the the coupling of the spherical harmonic compared to the plane wave ($g_{1,2}=124 \ \mu eV$) is resonantly enhanced:
\begin{equation}
\label{EQ:5}
\frac{g_{1,39}}{g_{1,2}}=-i 0.06 b_{1,39}\left({k r_0}\right) A^{1,2}_{1,39}\left({r_0 + \delta r}\right)   
\end{equation}
Here we utilized the fact that $B^{1,2}_{1,39} \ll A^{1,2}_{1,39}$.
While the resonant enhancement is provided by the $b_{1,39}$ Mie coefficient here, the translational coefficient reduces the effect. That is why if one tries to couple the evanescent light to the dipole transition the effect is much weaker as $A^{0,1}_{1,39} \ll A^{1,2}_{1,39}$. The resulting exciton - evanescent light coupling is shown in the Fig.\ref{FIG:1}    
\par
The coupling grows with mode number $l$, because the gradient of the evanescent field increases. At $l=89$ the coupling becomes of the order of $meV$. The property of the scalable coupling factor can be utilized in practical applications such as non-linear optics and is the subject of our future work.
\section{results and discussion} 
Around the resonance between WGM  and the quadrupole exciton $\omega_{1l} \approx \omega_{1S}$ the EP branches are given by the eigenvalues of the following Hamiltonian:
\begin{equation}
\label{EQ:6}
H / \hbar = \omega_{1l} a^\dag_k a_k + \omega_{1S} b^\dag_k b_k + \\
 g_{1l} \left({a^\dag_k b_k + a_k b^\dag_k}\right),
\end{equation}
where $a_k, \ b_k$ are annihilation operators for light and the exciton, respectively. 
Therefore, considering that both the exciton and WGM of a single sphere are localized, the dispersion is reduced to:
\begin{equation}
\label{EQ:7}
\omega = \omega_{1S} \pm g_{1l}/\hbar
\end{equation}
The excitons are trapped in the minimum of the lower branch. Hence the corresponding WGM pattern can be observed. The dispersion above is similar to the quadrupole-dipole hybrid in the organic-inorganic hetero-structures \cite{ROSLYAK:2007}. In the later case, the excited organic molecules create an evanescent field penetrating in the cuprous oxide. 
\par
\begin{figure}[htbp]
	\centering
	  \includegraphics[width=8cm]{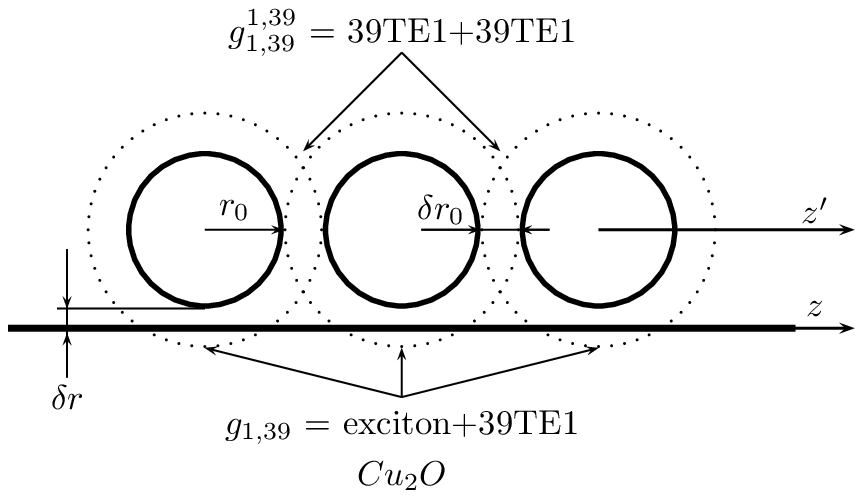}
	\caption{Schematic of formation of the evanescent polariton on linear chain of PMS. Actual dispersion is determined by the ratio of two coupling parameters such as exciton-WGM coupling and WGM-WGM coupling between the microspheres.}
	\label{FIG:2}
\end{figure}

The evanescent polariton provided by a single sphere gives the time coherence necessary for the observable BEC of the quadrupole exciton. But the spatial coherence is limited to a small region nearby to the sphere. To improve the spatial coherence one has to sacrifice the temporal coherence slightly by delocalizing the corresponding WGM. It can be done by using an array of spheres aligned along the $z$ direction and separated by the distance $\delta r_0$ (See Fig.\ref{FIG:2}).
\par
\begin{figure}[htbp]
	\centering
		\includegraphics[width=8.6cm]{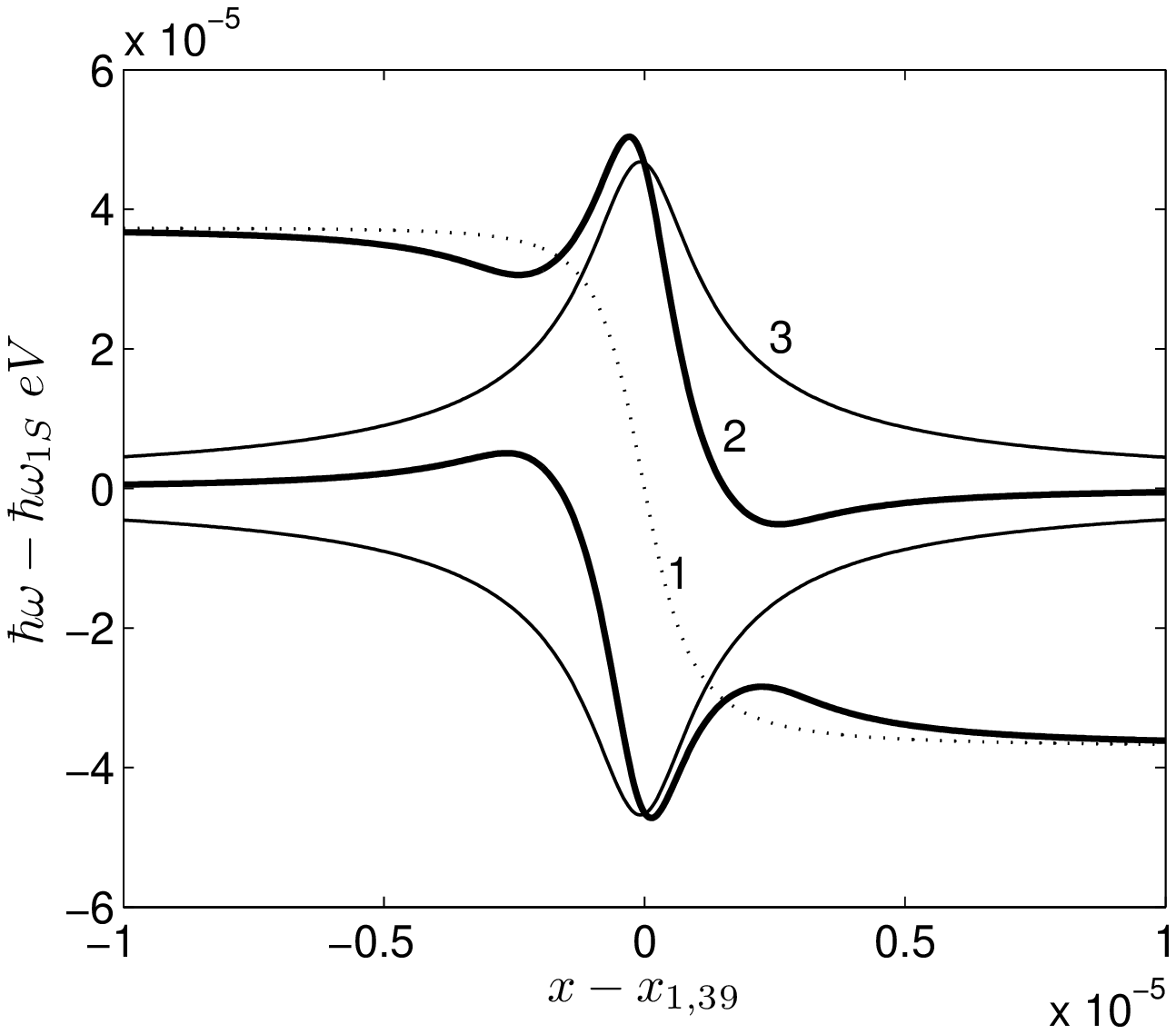}
	\caption{Dispersion of the evanescent polariton 39TE1. The dashed line (1) corresponds to the dispersion of the chain of spheres touching each other ($\delta r_0 = 0$). The thin solid line (3) stands for upper and lower branches of a single sphere dispersion ($\delta r_0 \gg \delta r = 0$). The thick solid curve (2) is the case of linear chain of the spheres in contact with the cuprous oxide ($\delta r_0 = \delta r = 0$).}
	\label{FIG:3}
\end{figure}

Recent experimental \cite{HARA:2005} and theoretical \cite{DEYCH:2006} studies have shown that the WGM can travel along the chain as "heavy photons". Therefore the WGM acquires the spatial dispersion, and the evanescent polariton has the form (See Fig.\ref{FIG:3}):
\begin{eqnarray}
\nonumber
2\omega = \omega_{1l,k}+\omega_{1S} \pm \sqrt{\left({\omega_{1l,k}-\omega_{1S}}\right)^2+4{\left|{g_{1l}/\hbar}\right|}^2}\\
\omega_{1l,k} = \omega_{1S} + 2 \left({g_{1l}^{1l}/\hbar}\right) cos(x-x_{1l}+\pi/2)  
\end{eqnarray}
Here $g_{1l}^{1l} = \omega_{1S} b_{1l} A_{1l}^{1l}\left({\delta r_1}\right)$ is the nearest-neighbor inter-sphere coupling parameter.
\par
When the coupling between spheres dominates ($\delta r \gg \delta r_0$) the minimum of the lower polariton branch disappears. Consequently, for BEC of the evanescent polariton one has to keep the desired balance between spatial and temporal coherence by adjusting experimental parameters $\delta r$ and $\delta r_0$.
\par
Both, the energy of the $1S$ quadrupole exciton and the WGM depend on the temperature. Therefore one can use a standard temperature scan to reveal the evanescent polariton dispersion \cite{PETER:2005}. 
\par
In summary, we emphasize that because of high localization of the quadrupole exciton they couple "naturally" to the WGM. Although the coupling of the evanescent field to the dipole allowed excitation are less pronounced then quadrupole ($\mathbf{d} \cdot \mathbf{E}_i \ll Q_{ij} \nabla_i E_{s,j}$) it can be greatly improved by trapping the excitons in low dimensional structures such as quantum wells or quantum dots \cite{FAN:1999}. 
\par
The theory developed above is applicable also for void cavities, spherical impurities and metallic droplets in bulk cuprous oxide crystal.     
\begin{acknowledgments}
We would like to acknowledge Ms. Upali Aparajita for useful comments and help with the manuscript. This work was supported in part by PCS-CUNY.  
\end{acknowledgments}
\section{Appendix}
In the appendix we list explicit expression for the Mie scattering coefficient:
\begin{eqnarray}
\nonumber
a_{ml} &=&\dfrac{n^2 j_{ml} \left(  nx\right)  \left[
xj_{ml} \left( x\right)  \right]  ^{\prime} -j_{ml} \left(
x\right) \left[ nxj_{ml} \left( nx\right)  \right]  ^{\prime}
}{n^{2} j_{ml} \left( nx\right)  \left[ xh_{ml}^{\left(  1\right)
} \left( x\right) \right]  ^{\prime} -h_{ml}^{\left(  1\right)  }
\left( x\right) \left[  nxj_{ml} \left( nx\right)  \right]
^{\prime} }\\
\nonumber
b_{ml}&=&\dfrac{j_{ml}\left(  nx\right)  \left[ xj_{ml}\left(
x\right)  \right]  ^{\prime}-n^{2}j_{ml}\left( x\right)  \left[
nxj_{ml}\left(  nx\right)  \right] ^{\prime}}{j_{ml}\left(
nx\right) \left[
xh_{ml}^{\left(  1\right)  }\left(  x\right)  \right]  ^{\prime}-n^{2} h_{ml}^{\left(  1\right)  }\left(  x\right)  \left[ nxj_{ml}\left(
nx\right) \right]  ^{\prime}}
\end{eqnarray}
Here $n = \epsilon^2$ is the refractive index of the spheres; $x = k r_0$ is the size parameter; $j_{ml}, \ h_{ml}$ are the spherical Bessel and Hankel of the first kind functions respectively.
\par
In the case of $l\gg 1$ the calculation of the translational coefficients
can be significantly simplified with the help of the so-called 
maximum term approximation\cite{MIYAZAKI:2000}. 
\begin{eqnarray*}
\nonumber
A_{l}^{l^\prime} & \cong&-2l\left(  -1\right)
^{l+1} \sqrt{\dfrac{l+l^\prime}{\pi\left(  l^\prime+1\right)  \left(
l-1\right) } } \times \\
& & \dfrac{l^{l} {(l^{\prime})}^{l^\prime}
}{\left(l^\prime+1\right)^{l^\prime+1} \left(
l-1\right)^{l-1} }h_{l+l^\prime}^{(1)} \left(  \eta x \right)\\
B_{l}^{l{^\prime}} & \cong &i\dfrac{x\left|
i-j\right|}{ll^\prime}A_{l}^{l{^\prime}}
\end{eqnarray*}
Here $\eta$ defined as $\eta=|r_0+\delta r|/r_0\ge 1$ is
a dimensionless distance between the centers of the spheres.
 
\bibliography{ev_polariton} 
\end{document}